\documentstyle[12pt,cite,shadow,epsf,epsfig,psfrag]{article}
\begin{document}
\begin{flushright}
preprint UT-778  \\
December, 1997 \\
\end{flushright}

\newcommand{\be}{\begin{equation}}
\newcommand{\ee}{\end{equation}}
\newcommand{\bea}{\begin{eqnarray}}
\newcommand{\eea}{\end{eqnarray}}
\newcommand{\La}{\Lambda}
\newcommand{\no}{\nonumber}

\begin{center}
{\Large \bf Relation between  S-duality in N=2 SQCD \\
           and \\
 Non-Abelian Duality in N=1 SQCD \\
         }
\end{center}

\begin{center}
\medskip
Jun Furukawa\\
\medskip
{\it  Department of Physics, University of Tokyo \\
Bunkyo-ku, Tokyo 113,Japan  \\
\medskip }
furukawa@hep-th.phys.s.u-tokyo.ac.jp
\end{center}
\bigskip
\bigskip

\begin{abstract}
With the help of M-theory configuration of N=1 supersymmetric QCD, we analyze the strong coupling moduli of N=1 theory and its relation with S-duality transformation in N=2 supersymmetric QCD. As a result we confirm that two type-IIA descriptions for N=1 supersymmetric QCD, one of them being strong coupling description and another being weak, correspond to a single M-theory description for N=1 supersymmetric QCD. The existence of singlet fields is also discussed. 
\end{abstract}

\newpage

\noindent{\bf Introduction}

A remarkable progress has been made in the last few years in understanding of supersymmetric gauge theory in four dimensions from both gauge theory itself and effective world volume theory of brane\cite{BY1}. Non-Abelian duality in N=1 supersymmetric gauge theory\cite{L5} is the most interesting among them and at the same time is still mysterious. This duality has not been fully understood from the point of view of field theory, although exact results are known in N=2 supersymmetric QCD\cite{C8} and several approaches to understand N=1 theory from N=2 theory are proposed\cite{LMS10}. On the other hand, understanding of non-Abelian duality has made progress from the method of branes. It is now considered that the duality corresponds to the exchange of two NS-5 branes in type IIA string theory\cite{B4}. Further, it was shown that the exact solution of the N=2 super symmetry QCD can be seen as the configuration of M-theory five brane embedded in the Taub-NUT space\cite{W3}, and the M-theory configuration of N=1 supersymmetric QCD is also derived\cite{BY2}. 

Our starting point in this article is the N=2 supersymmetric $SU(N_c)$ gauge theory with $2N_c$ flavors\cite{C7} which is known to be finite. By adding a mass term for the adjoint chiral field and for $2N_c-N_f$ flavors, we obtain the N=1 supersymmetric $SU(N_c)$ gauge theory with $N_f$ flavors. Adding the mass term for the adjoint field to N=2 theory is equivalent to rotating\cite{B11} the brane at baryonic branch root in M-theory picture. Embedding in the finite theory makes it possible to analyze strong coupling behavior of the N=1 theory.

\bigskip 
\noindent{\bf S-duality and Baryonic branch root of massless finite theory}
\\

We start our study by considering N=2 supersymmetric $SU(N_c)$ QCD with $2N_c$ flavors. The curve for this finite theory is given by   
\bea
t^2-\prod_{i=1}^{N_c}(v-\phi_i)~t+{(1-g(\tau)^2) \over 4} \prod_{i=1}^{2N_c}(v+g(\tau)m+m_{i})=0
\label{finite}
\eea
where
\bea
\sum_i^{N_c} \phi_i &=& 0        \no \\
m   &=& {1 \over 2N_c}\sum_{i=1}^{2N_c}M_i  \no \\
m_i &=& M_i-m                        \no \\
g(\tau)      &\equiv & \frac{\theta_2^4(\tau)+\theta_1^4(\tau)}{\theta_2^4(\tau)-\theta_1^4(\tau)}
  ~~~~~~~or~~~~~~\frac{\theta_3^4(\tau)-\theta_1^4(\tau)}{\theta_3^4(\tau)+\theta_1^4(\tau)}
\eea
The parameters $M_i$'s are the diagonalized bare masses for the $2N_c$ hyper multiplets. The parameters $\phi$'s are the diagonalized vevs of the lowest components of the adjoint chiral super fields. $g(\tau)$ is the function of coupling $\tau$ and defined as in \cite{C7}. It is now known that this curve corresponds to the configuration of 5-branes in M-theory. Above parameters are interpreted as the following complex coordinates,  $v=x^4+ix^5, s=(x^6+ix^{10})/ R, t=\exp{(-s)}$ where $x^{10}$ is the eleventh coordinate of M-theory which is compactified on a circle of radius $R$. 

Now we consider the massless case $M_i=0$ and rotate the N=2 theory curve (\ref{finite}) with and reduce it to that of N=1 theory. Moduli space of the N=2 Coulomb branch will collapse into the singuralities when perturbed by the mass term for the adjoint field, and they all coincide with the baryonic branch root when appropriate limits are taken\cite{BY2}. The curve at the baryonic branch root of massless N=2 theory is as follow.
\bea
t^2-\left ({1-g \over 2}v^{N_c}+{1+g \over 2}v^{N_c}\right ){}t+{1-g^2 \over 4}v^{2N_c}=0    
\label{masslessN2}
\eea
The curve at this special point in moduli space is determined by the condition that it must be factorized. The coefficient of term $v^{N_c}$ is one and the coefficient of term $v^{N_c-1}$ is zero. 

Now by introducing the complex coordinate $w=x^8+ix^9$, we rotate the brane configuration of N=2 (\ref{masslessN2}) to N=1 and obtain a curve 
\bea
\left (\tilde{t}-{1-g \over 2}\mu^{N_c}v^{N_c}\right ) \left (\tilde{t}-{1+g \over 2}w^{N_c}\right )=0 
\label{a}
\eea
where $\tilde{t}=\mu^{N_c}t, \omega = \mu v$, with $\mu$ being the mass of the adjoint field. And ${1-g \over 2}\mu^{N_c}=\Lambda^{N_c}_{N=1}\equiv{1-g^2 \over 4}\mu^{N_c}$ at the limit of decoupling.


Now we change the value of coupling from weak to strong as $g \rightarrow -g$, that is, $ \tau \rightarrow -{1 \over \tau}$ within the baryonic branch root. The curve obtained 
\bea
\left (\tilde{t}-{1+g \over 2}\mu^{N_c}v^{N_c}\right ) \left (\tilde{t}-{1-g \over 2}w^{N_c}\right )=0 
\eea
is exactly the same as the equation(\ref{a}) since $w=\mu v$. This is a manifestation of the S-duality in the massless finite theory.

\bigskip 
\noindent{\bf Non-Abelian duality and Baryonic branch root of $N_f(<2N_c)$ flavor theory}
\\

The N=2 theory with arbitrary $N_f(<2N_c)$ massless flavors is obtained by decoupling the $k=2N_c-N_f$ flavors from the finite theory. For convenience we choose the values of masses of finite N=2 theory to be
\bea
M_n =\left \{
\begin{array}{ll}
  M~e^{2 \pi i n \over k}  &(n=1,\dots,k) \\
    0                    & (n=others)
\end{array}\right .
\label{mass}
\eea
The curve for the baryonic branch root of this theory becomes
\bea
t^2-\left [{1-g \over 2 }v^{N_f-N_c}(v^k-M^k)+{1+g \over 2}v^{N_c} \right ]~t+{1-g^2 \over 4}v^{N_f}(v^k-M^k)=0
\label{N2broot}
\eea
This corresponds to a choice of $\phi$'s 
\bea
\phi_n =\left \{
\begin{array}{ll}
|({1-g \over 2})^{1 \over k}M|e^{2 \pi i n \over k}\simeq \Lambda_{N=2,N_f}&(n=1,\dots,k)\\
0 & n=others
\end{array}\right .
\label{phiweak}
\eea
\begin{figure}[htb]
\begin{center}
\epsfxsize=3in\leavevmode\epsfbox{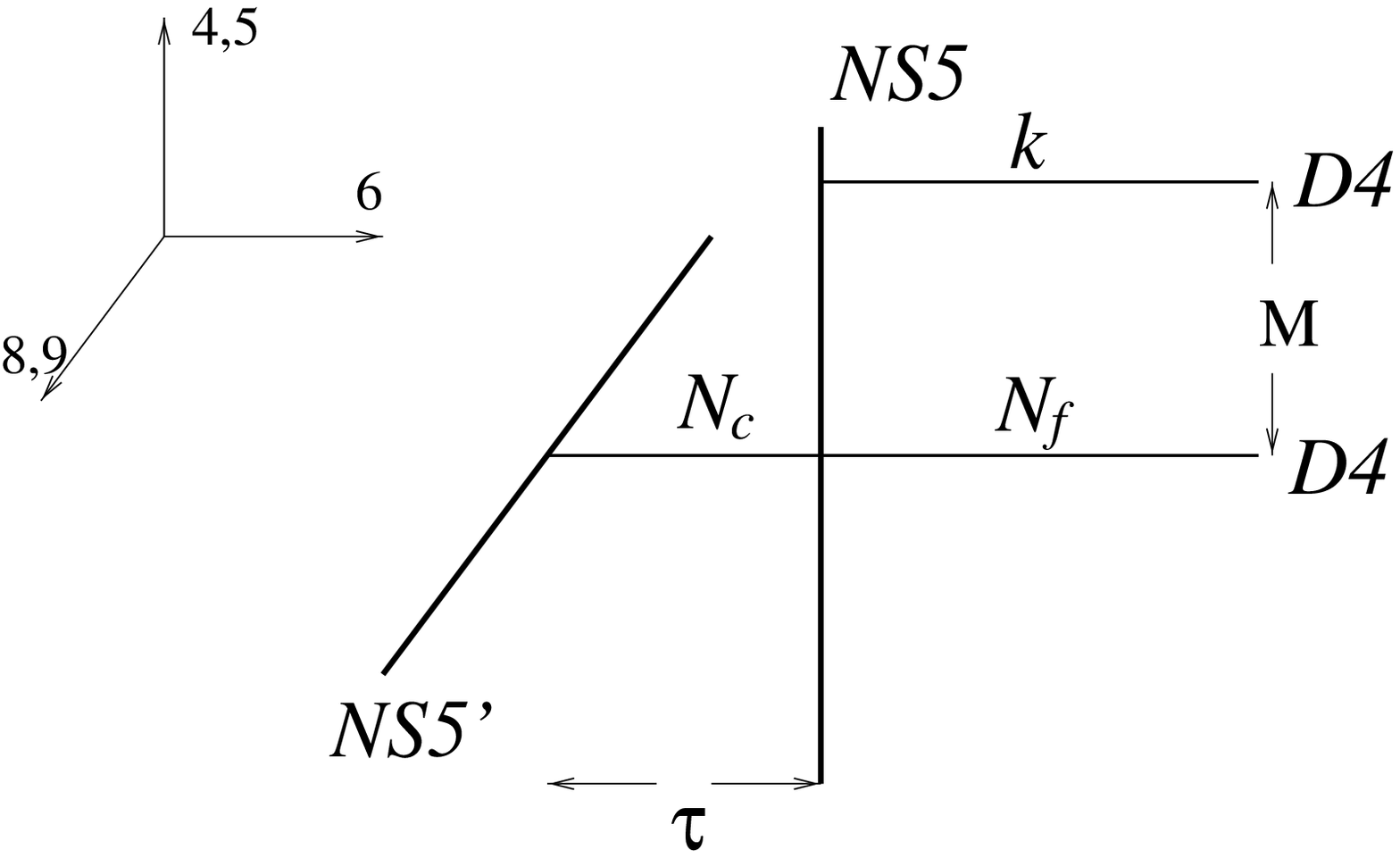}
\caption{Weak coupling picture}
\label{fig1}
\epsfxsize=3in\leavevmode\epsfbox{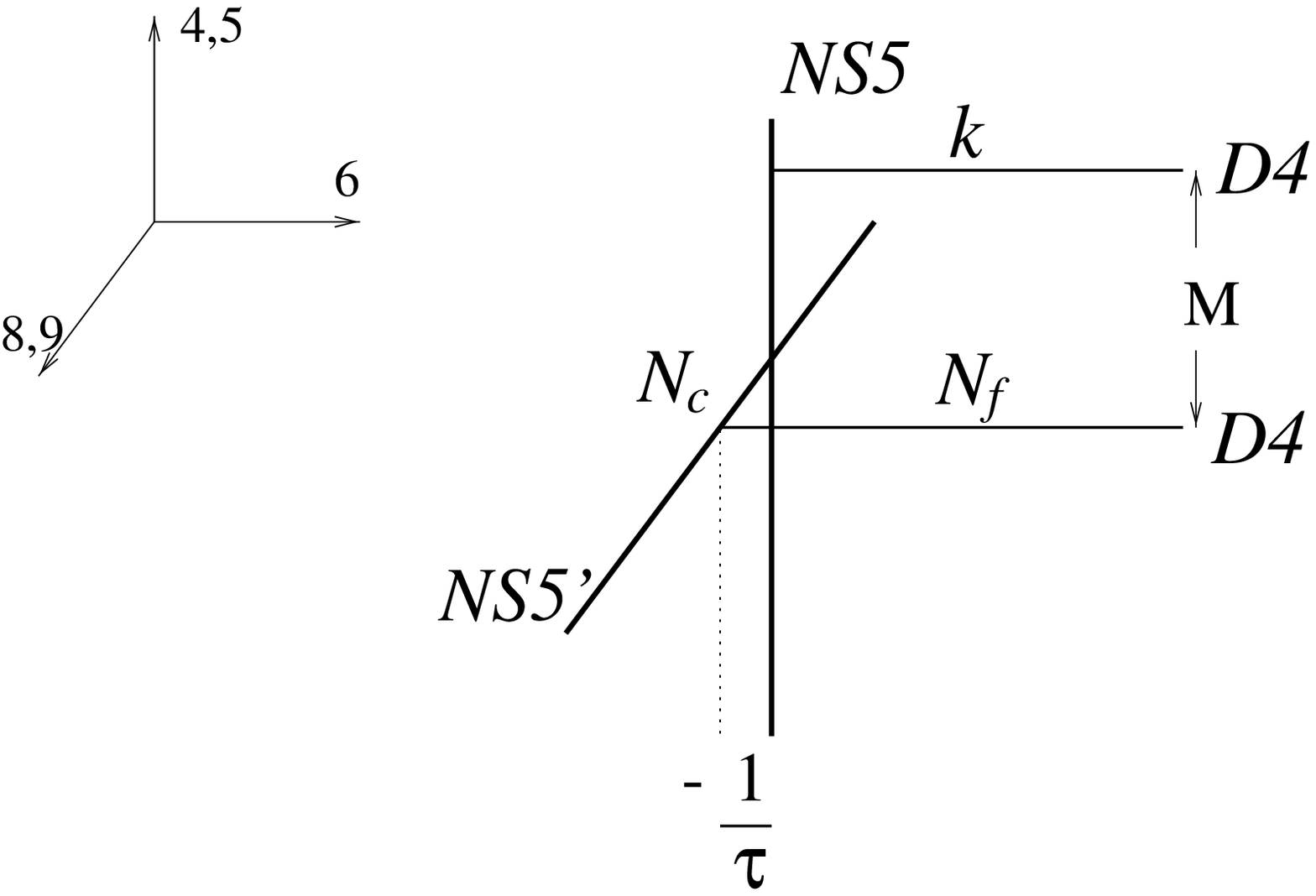}
\end{center}
\caption{Strong coupling picture}
\label{fig2}
\end{figure}
\begin{figure}[htb]
\begin{center}
\epsfxsize=3in\leavevmode\epsfbox{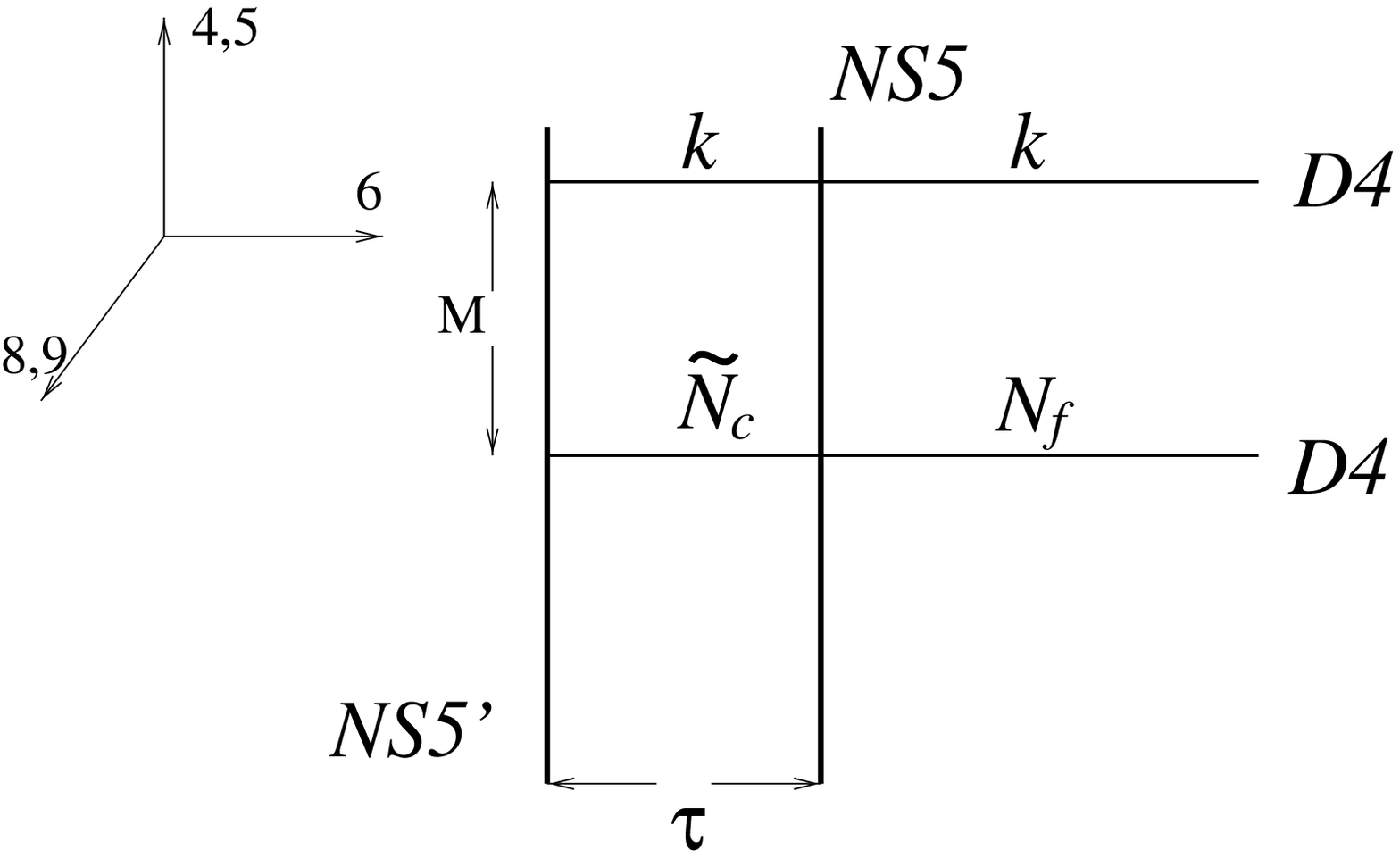}
\caption{N=2 weak coupling picture}
\label{fig3}
\epsfxsize=3in\leavevmode\epsfbox{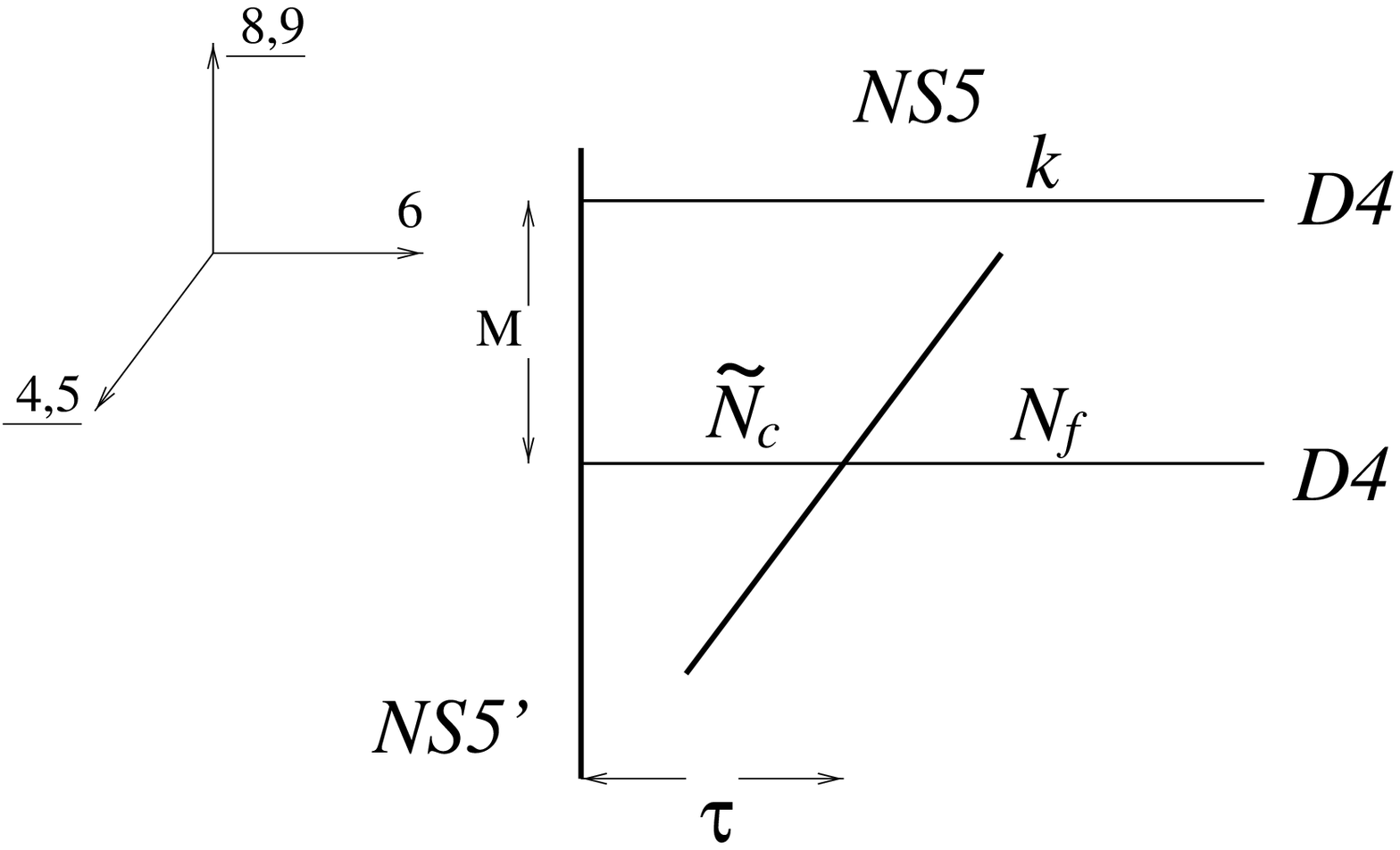}
\end{center}
\caption{N=1 weak coupling picture}
\label{fig4}
\end{figure}
This agrees with the known baryonic branch root of $N_f(<2N_c)$ flavor theory at the limit of $M \rightarrow \infty$ and $(1-g) \rightarrow 0$ keeping $\Lambda_{N=2,N_f}^k={1-g^2 \over 4}M^k$ fixed. 

We factorize the curve (\ref{N2broot}) and rotate it as before and obtain
\bea
\left ( \tilde{t}+{1-g \over 2}M^k~\mu^{N_c}~v^{N_f-N_c}-{1-g \over 2}\mu^{N_c}v^{N_c} \right )
\left ( \tilde{t}-{1+g \over 2}\omega^{N_c}\right )=0 .
\label{eq1}
\eea
At the limit of $g \rightarrow 1$ and $\mu \rightarrow \infty$ keeping $\Lambda^{3N_c-N-f}_{N=1}\equiv {(1-g^2) \over 4}\mu^{N_c}M^k$ fixed, $k$ flavors are decoupled. Thus we obtain
\bea
\left ( \tilde{t}+\Lambda^{3N_c-N_f}_{N=1}~v^{N_f-N_c} \right )\left ( \tilde{t}-w^{N_c}\right )&=&0 
\eea

Now by applying the transformation $g \rightarrow -g, \tau \rightarrow -{1 \over \tau}$ to the curve (\ref{eq1}) we obtain a strong coupling description. 
\bea
\left ( \tilde{t}+{1+g \over 2}M^k~\mu^{N_c}~v^{N_f-N_c}-{1+g \over 2}\mu^{N_c}v^{N_c} \right )
\left ( \tilde{t}-{1-g \over 2}\omega^{N_c}\right )=0
\label{N1stlimit} 
\eea
By introducing new parameters $\bar{t}=t \tilde{\mu}^{\tilde{N_c}}/M^k, \tilde{w}=\tilde{\mu}v, \tilde{\mu}=-\mu$, one may rewrite (\ref{N1stlimit}) as 
\bea
\left ( \bar{t}+{1+g \over 2}\tilde{w}^{\tilde{N_c}}-{1+g \over 2}{\tilde{w}^{N_c}\over M^k \tilde{\mu}^k} \right )
\left ( \bar{t}-{1-g \over 2}{\tilde{\mu}^{\tilde{N_c}} \over M^k} v^{N_f-\tilde{N_c}}\right )=0
\label{eq2}
\eea
Further, by defining $\tilde{\Lambda}_{N=1}$ as $ \tilde{\Lambda}^{3\tilde{N_c}-N_f}_{N=1}=(1-g^2)\tilde{\mu}^{\tilde{N_c}}/4 M^k$, the curve at the decoupling limit of strong coupling will be
\bea
\left (\bar{t}+\tilde{w}^{\tilde{N_c}} \right ) \left (\bar{t}-\tilde{\Lambda}^{3\tilde{N_c}-N_f}_{N=1} v^{N_f-\tilde{N_c}} \right )=0
\eea
The charges of above parameters are
\bea
\begin{array}{cccl}
&U(1)_{4,5}&U(1)_{8,9}&\\
v, M&2&0&\\
w, \tilde{w}&0&2&\\
\mu, \tilde\mu&-2 & 2 \\
\tilde{t}& 0&2 N_c&\\
\bar{t}&0&2 \tilde{N_c}&\\
\Lambda^{3N_c-N_f}_{N=1}&2N_c-2N_f&2 N_c&\\
\tilde{\Lambda}^{3\tilde{N}_c-N_f}_{N=1}&2\tilde{N}_c-2N_f&2 \tilde{N}_c&\\
\end{array}
\eea
Note that $N_c$ is replaced by $\tilde{N_c}$ going from $\tilde{t},\Lambda_{N=1}$ to $\bar{t},\tilde{\Lambda}_{N=1}$.

This N=1 configuration of strong coupling (\ref{N1stlimit}) can be obtained by rotating following N=2 configuration.
\bea
\left (t+{1+g \over 2}M^k~v^{N_f-N_c}-{1+g \over 2}v^{N_c} \right )
\left ( t-{1-g \over 2}v^{N_c}\right )=0 
\label{eq3}
\eea
This corresponds to a choice of $\phi$'s
\bea
\phi_n =\left \{
\begin{array}{ll}
|({1+g \over 2})^{1 \over k}M|e^{2 \pi i n \over k}\simeq M&(n=1,\dots,k)\\
0 & (n=others)
\end{array}\right .
\label{eq4}
\eea
instead of (\ref{phiweak}). Therefore the curve (\ref{eq3}) represents N=2 weak coupling theory with $N_f$ flavors and $N_f-N_c$ colors at the decoupling limit $M \rightarrow \infty$.

This N=1 configuration (\ref{N1stlimit}) represents the strong coupling theory of N=1 supersymmetric $SU(N_c)$ QCD with $N_f$ flavors by construction, at the same time it can be interpreted as the weak coupling theory of N=1 supersymmetric $SU(N_f-N_c)$ QCD with $N_f$ flavors with singlet $M^i_j$ interacting with the quarks $q_i$'s by the superpotential $q M \tilde{q}$. The hypermultiplets $q_i$'s transform in the fundamental representation of $SU(N_f-N_c)$.

This can be seen by the following type-IIA pictures. Figure(\ref{fig1}) shows weak coupling theory of equation (\ref{eq1}). $N_c$ $D4$ branes are suspended between two $NS5$ branes. $N_f$ semi-infinite $D4$ branes which correspond to massless matter touch $NS5$ brane at the same place as $N_C D4$ branes. Each of $k$ semi-infinite $D4$ branes corresponding to massive matter touches $NS5$ brane at $v=M~e^{2 \pi i n \over k}$. However, in the following pictures, they are drawn at the same place for simplification. Its strong coupling configuration equation (\ref{eq2}) will be figure(\ref{fig2}) by construction. Only the value of $\tau$ is changed to $-{1 \over \tau}$ from figure(\ref{fig1}). 

On the other hand, considering equation(\ref{eq4}), N=2 configuration equation(\ref{eq3}) can be understood as to represents the $SU(N_f-N_c=\tilde{N}_c)$ weak coupling $\tau$ gauge theory with $N_f$ flavors which is obtained from $N_c$ color and $2N_c$ flavor theory by decoupling $k$ colors and flavors as figure(\ref{fig3}). Therefore, equation(\ref{eq2}) which is the rotated curve of equation(\ref{eq3}) can be interpreted as N=1 weak coupling theory of $SU(N_f-N_c)$ gauge group as shown in figure(\ref{fig4}). 

Comparing two type-IIA pictures figure(\ref{fig2}) and figure(\ref{fig4}), we see that different two type-IIA pictures, one is weak and the other is strong coupling, corresponds to a single M-theory configuration equation(\ref{eq2}). In the M-theory N=1 configuration, two M5 branes which was NS5 branes in type-IIA cross each other. But in type IIA picture they must be parallel and separately connected each other by suspended $D4$ branes. Two pictures emerge when we decide which of two branes comes left or right in type-IIA picture. This is the Non-Abelian duality from the M-theoretical point of view.

\bigskip 
\noindent{\bf Discussions}

In the limit of $\mu \rightarrow \infty$ and weak coupling of $(1-g) \rightarrow 0$ or strong coupling of $(1+g) \rightarrow 0$ keeping ${1-g^2 \over 4} \mu^{N_c} M^k=\Lambda^{3N_c-N_f}_{N=1}$ fixed, theories are supposed to describe $N=1$ supersymmetric QCD. The relation between the strong coupling scale $\Lambda_{N=1}$ and the scale of its dual $\tilde{\Lambda}_{N=1}$ are
\bea
\Lambda_{N=1}^{3N_c-N_f}~\tilde{\Lambda}_{N=1}^{3\tilde{N}_c-N_f}
&&=\left ({1-g^2 \over 4}M^k~\mu^{N_c} \right )
 \left ({1-g^2 \over 4}{\tilde{\mu}^{\tilde{N_c}}\over M^k} \right ) \no \\
&&= {(1-g^2)^2 \over 16 } \mu^{N_f}(-1)^{\tilde{N}_c} \no  \\
&&\equiv  \bar{\mu}^{N_f}(-1)^{\tilde{N_c}}
\label{eq5}
\eea
This is similar to the equation which Seiberg has proposed\cite{L5}. But it is not possible for the both scales $\Lambda_{N=1}$ and $\tilde{\Lambda}_{N=1}$ to take non-zero finite value simultaneously. Further, the value of $\bar{\mu}$ is not determined. The is a limitation of our analysis. 

We have been considering only the case of mass given by equation (\ref{mass}). However, arbitrary large mass limit of decoupling hypermultiplets should be allowed. Therefore, we consider the case of $m={1 \over 2 N_c}\sum^{2N_c}_{i=1}M_i \neq 0$ instead of (\ref{mass}). Candidate for the baryonic branch of the curve with nonzero $m$ is
\bea
t^2-\left ({1-g \over 2}V^{N_f-N_c} \prod^k_{i=1}(V+M_i)+{1+g \over 2}V^{N_c}\right )~t+{1-g^2 \over 4}V^{N_f} \prod^k_{i=1}(V+M_i)=0
\eea
where $V=v+(1-g)m$. The condition $\sum \phi_i =0$ are satisfied only in the decoupling limit of $(1-g) \rightarrow 0$ and consistent with the above argument.

The existence of singlets is understood as follow. Consider the $N=2$ theory with the product gauge group $SU(N_c) \otimes SU(N_f=2N_c)$, whose super potential is
\bea
W=\sqrt{2}Q_i \Phi \tilde{Q}^i+\sqrt{2}Q_i M^i_j \tilde{Q}^j
\eea
where $\Phi$ and $M$ are adjoint chiral fields of the gauge group $SU(N_c)$ and $ SU(N_f)$ respectively. The hypermultiplets $Q$'s transform in obvious notation, as ($N_c,\bar{N_f}$). The curve for the theory is
\bea
t^3-\prod_{i=1}^{N_c}(v-\phi_i)~t^2+{(1-g^2) \over 4} \prod_{i=1}^{2N_c}(v+g\mu+\mu_{i})~t+\Lambda^{3N_c}_{g,N=2}=0
\label{gauged}
\eea

The original theory is obtained at the weak coupling limit of $\Lambda_g=0$ which means to vanish the kinetic term of $M$. The same conclusion as (\ref{eq1}) is derived from the $N=1$ curve
\bea
\left ( \tilde{t}+{1-g \over 2}M^k~\mu^{N_c}~v^{N_f-N_c}-{1-g \over 2}\mu^{N_c}v^{N_c} \right )
\left ( \tilde{t}-{1+g \over 2}\omega^{N_c}\right )~\tilde{t}
+\Lambda_{g,N=1}^{3N_c}=0 
\eea
which is the rotated curve of N=2 curve(\ref{gauged}). We choose the values of $\phi$'s so as to coincide with the curve at baryonic branch root at the limit of $\Lambda_{g,N=1}\rightarrow 0$. Same $N=1$ curve (\ref{eq1}) and decoupled curve $\tilde{t}=0$ are obtained at the limit of $\Lambda_{g,N=1} \equiv \mu \Lambda_{g,N=2} \rightarrow 0$. On the other hand, from the strong coupling curve of (\ref{gauged}), we obtain a curve,
\bea
 \left ( \bar{t}+{1+g \over 2}\tilde{w}^{\tilde{N_c}}-{1+g \over 2}{\tilde{w}^{N_c}\over M^k \tilde{\mu}^k} \right )
\left ( \bar{t}-{1-g \over 2}{\tilde{\mu}^{\tilde{N_c}} \over M^k} v^{N_f-\tilde{N_c}}\right )~\bar{t}+{\Lambda_{g,N=1}^{3N_c} \over \mu^{3k}M^{3k}}=0
\eea
Thus the same $N=1$ curve (\ref{eq2}) and decoupled curve $\bar{t}=0$ are obtained without taking the limit of $\Lambda_{g,N=1}\rightarrow 0$ when $2N_c > N_f$, which suggests the existence of singlet fields in the dual description.

\end{document}